\documentclass[useAMS,usenatbib]{mn2e}

\usepackage{natbib}
\usepackage{graphicx}
\usepackage{amssymb}
\usepackage{color}
\usepackage{caption}

\title[Radiative feedback in dusty quasars]{AGN radiative feedback in dusty quasar populations}

\author[ ]
{W. Ishibashi$^{1}$\thanks{E-mail:
wako.ishibashi@physik.uzh.ch}, M. Banerji$^{2,}$$^{3}$ and A. C. Fabian$^{2}$
\footnotemark[0]\\
$^{1}$Physik-Institut, Universitat Zurich, Winterthurerstrasse 190, 8057 Zurich, Switzerland 
\footnotemark[0]\\
$^{2}$Institute of Astronomy, Madingley Road, Cambridge CB3 0HA \\
$^{3}$Kavli Institute of Cosmology Cambridge, Madingley Road, Cambridge CB3 0HA, UK
}

\begin{document}

\pdfminorversion=4

\date{Accepted ? Received ?; in original form ? }

\pagerange{\pageref{firstpage}--\pageref{lastpage}} \pubyear{2012}

\maketitle

\label{firstpage}

\begin{abstract} 

New populations of hyper-luminous, dust-obscured quasars have been recently discovered around the peak epoch of galaxy formation ($z \sim 2-3$), in addition to similar sources found at lower redshifts. Such dusty quasars are often interpreted as sources `in transition', from dust-enshrouded starbursts to unobscured luminous quasars, along the evolutionary sequence. Here we consider the role of the active galactic nucleus (AGN) radiative feedback, driven by radiation pressure on dust, in high-luminosity, dust-obscured sources. We analyse how the radiation pressure-driven dusty shell models, with different shell mass configurations, may be applied to the different populations of dusty quasars reported in recent observations. We find that expanding shells, sweeping up matter from the surrounding environment, may account for prolonged obscuration in dusty quasars, e.g. for a central luminosity of $L \sim 10^{47}$erg/s, a typical obscured phase (with extinction in the range $A_{V} \sim 1-10$ mags) may last a few $\sim 10^6$yr. On the other hand, fixed-mass shells, coupled with high dust-to-gas ratios, may explain the extreme outflows recently discovered in red quasars at high redshifts. We discuss how the interaction between AGN radiative feedback and the ambient medium at different temporal stages in the evolutionary sequence may contribute to shape the observational appearance of dusty quasar populations. 
\end{abstract}



\section{Introduction}

Observations over the past few years have uncovered new populations of high-luminosity, dust-reddened quasars at $z \sim 2-3$, when both active galactic nuclei (AGN) and star formation activities are known to peak. 
Near-infrared selection has identified a population of hyper-luminous ($L \sim 10^{47}$erg/s), dusty quasars with significant dust extinction ($A_{\rm V} \sim 2-6$ mag), comparable to that of submillimetre galaxies \citep{Banerji_et_2012, Banerji_et_2015}. 
A unique population of extremely red quasars (ERQs), with high luminosities ($L \sim 10^{47}$erg/s) and peculiar emission line features, has also been recently discovered in BOSS. \citep{Zakamska_et_2016, Hamann_et_2017}. 
Another class of hyper-luminous, dust-obscured galaxies, known as Hot DOGs, has been previously unveiled by WISE \citep{Tsai_et_2015, Assef_et_2015}. These intrinsically luminous ($L \sim 10^{47}-10^{48}$erg/s) sources are characterised by heavy extinction (with $A_{\rm V} \sim 50$ mag). X-ray observations indicate that Hot DOGs are powered by heavily obscured, possibly Compton-thick, AGNs \citep{Stern_et_2014}.  

Such luminous, dust-obscured AGNs have often been interpreted as sources `in transition', from the dust-enshrouded starburst phase to the unobscured QSO stage, within the evolutionary scenario \citep[of e.g.][]{Sanders_et_1988}. In this picture, the dusty quasars, e.g. dust-reddened type I quasars \citep{Banerji_et_2015}, are likely observed in the short-lived blow-out phase, when feedback from the central black hole is expelling its dust cocoon. Indeed, some of the red quasars show direct evidence for outflows \citep{Banerji_et_2012, Zakamska_et_2016}, and a particularly high BAL fraction is found among the dust-reddened quasars and the ERQ population \citep{Urrutia_et_2009, Hamann_et_2017}.

The occurrence of high luminosities, coupled with the presence of significant amounts of dust, should form favourable conditions for AGN radiative feedback. The importance of radiation pressure on dust as a physical mechanism for driving AGN feedback has already been highlighted in the past \citep{Fabian_1999, Murray_et_2005}. More recently, \citet{Thompson_et_2015} have analysed the dynamics of dusty radiation pressure-driven shells, showing that the asymptotic velocities can exceed the escape velocities in different astrophysical sources. In the context of AGN feedback, we have further examined the role of radiation pressure on dust in driving powerful outflows on galactic scales \citep{Ishibashi_Fabian_2015}. We have also suggested AGN radiative feedback, which directly acts on the obscuring dusty gas, as a physical mechanism explaining the observed evolutionary sequence \citep{Ishibashi_Fabian_2016b}. 
Here we wish to analyse how our radiative feedback-driven shell models may be applied to the recently discovered populations of dusty quasars.


\section{Radiative feedback}
\label{Sect_radiative_feedback}

We consider AGN feedback driven by radiation pressure on dust. Radiation from the central source is absorbed by dust grains, and the ambient dusty gas is swept up into an outflowing shell. 
We assume a thin shell approximation subtending $4 \pi$ for simplicity \citep[cf][]{Thompson_et_2015}. 
We recall that the general form of the equation of motion is given by 
\begin{equation}
\frac{d}{dt} [M_{\rm sh}(r) v] = \frac{L}{c} (1 + \tau_{\rm IR} - e^{-\tau_{\rm UV}}) - \frac{G M(r) M_{\rm sh}(r)}{r^2}
\end{equation} 
where $L$ is the central luminosity, $M(r)$ the total mass distribution, and $M_{\rm sh}(r)$ the shell mass \citep[][]{Thompson_et_2015,  Ishibashi_Fabian_2015, Ishibashi_Fabian_2016b}. 
The infrared (IR) and ultraviolet (UV) optical depths are given by: 
\begin{equation}
\tau_{\rm IR,UV}(r) = \frac{\kappa_{\rm IR,UV} M_{\rm sh}(r)}{4 \pi r^2}
\end{equation}
where $\kappa_{\rm IR}$=$5 cm^2 g^{-1} f_{dg, MW}$ and $\kappa_{\rm UV}$=$10^3 cm^2 g^{-1} f_{dg, MW}$ are the IR and UV opacities, with the dust-to-gas ratio normalised to the Milky Way value. 
The shell column density is defined as 
\begin{equation}
N_{\rm sh}(r) = \frac{M_{\rm sh}(r)}{4 \pi m_p r^2}
\end{equation}


The importance of radiative feedback depends on the interaction with the ambient medium. 
We consider different assumptions for the shell mass configuration: fixed-mass shells with constant mass ($M_{\rm sh}(r) = M_{\rm sh}$), and expanding shells sweeping up matter from the surrounding environment. In the latter case, the density distribution of the ambient medium can be parametrized as a power law of radius, $r$, with slope $\alpha$:
\begin{equation}
n(r) = n_0 \left( \frac{r}{R_0} \right)^{-\alpha} \propto \frac{1}{r^{\alpha}}
\end{equation}
where $n_0$ is the density of the external medium and $R_0$ the initial radius. 
The corresponding swept-up shell mass is given by:
\begin{equation}
M_\mathrm{{sh}}(r) = 4 \pi m_p \int n(r) r^2 dr = 4 \pi m_p n_0 R_0^{\alpha} \frac{r^{3-\alpha}}{3-\alpha}
\end{equation}

In this section, we consider the simple case of the isothermal potential ($M(r) = \frac{2 \sigma^2}{G} r$, where $\sigma$ is the velocity dispersion), for which analytical limits can be derived. We first compare the fixed-mass shell and the $\alpha = 2$ shell model, which corresponds to an isothermal distribution (`isothermal shell' in the following). In the latter case, the ambient gas density is assumed to follow an isothermal distribution, which may be a reasonable approximation in the inner regions (on $\lesssim$ few kpc scales). The corresponding swept-up mass increases linearly with radius as $M_\mathrm{{sh}}(r) = 4 \pi m_p R_0^2 n_0 r$. Other forms of gravitational potential and shell mass configurations will be considered in the next section. 


\subsection{Effective Eddington ratio and shell dynamics}

The effective Eddington ratio is given by
\begin{equation}
\Gamma = \frac{L}{L_E'} = \frac{L r^2}{c G M(r) M_{\rm sh}(r)} (1 + \tau_{\rm IR} - e^{-\tau_{\rm UV}}) 
\end{equation}
where $L_E'$ is the effective Eddington luminosity, obtained by equating the outward force due to radiation pressure to the inward force due to gravity \citep{Ishibashi_Fabian_2016b}. 
We recall that there are three distinct physical regimes depending on the optical depth of the medium: optically thick to both UV and IR, optically thick to UV but optically thin to IR (single scattering limit), and optically thin to UV. In the IR-optically thick and UV-optically thin regimes, the Eddington ratio is independent of the shell mass configuration, and is only determined by the underlying potential:
\begin{equation}
\Gamma_{\rm IR,UV} = \frac{\kappa_{\rm IR,UV} L}{4 \pi G c M(r)} = \frac{\kappa_{\rm IR,UV} L}{8 \pi c \sigma^2 r}
\label{Gamma_IR_UV}
\end{equation}
Thus $\Gamma_{\rm IR}$ and $\Gamma_{UV}$ are the same for the fixed-mass and isothermal shell models, both decreasing with radius as $\Gamma_{\rm IR,UV} \propto 1/r$. 
In contrast, the Eddington ratio in the single scattering regime explicitly depends on the shell mass configuration:
\begin{equation}
\Gamma_{\rm SS} = \frac{L r^2}{c G M(r) M_{\rm sh}(r)} = \frac{L r}{2 c \sigma^2 M_{\rm sh}(r)}
\end{equation}
For fixed-mass shells:
\begin{equation}
\Gamma_{\rm SS,fix} = \frac{L r}{2 c \sigma^2 M_{\rm sh}}
\end{equation}
while for isothermal shells:
\begin{equation}
\Gamma_{\rm SS,iso} = \frac{L}{8 \pi m_p c \sigma^2 R_0^2 n_0}
\label{Gamma_SS_iso}
\end{equation} 

We see that in the case of fixed-mass shells, the Eddington ratio increases with radius (scaling as $\Gamma_{\rm SS,fix} \propto r$), and becomes increasingly super-Eddington \citep{Thompson_et_2015}; whereas for isothermal shells, the Eddington ratio is independent of radius and remains constant ($\Gamma_{\rm SS,iso} \sim \rm cst$). 

\begin{figure}
\centering
\textbf{\quad \quad Effective Eddington ratio vs. radius} 
\begin{center}
\includegraphics[angle=0,width=0.4\textwidth]{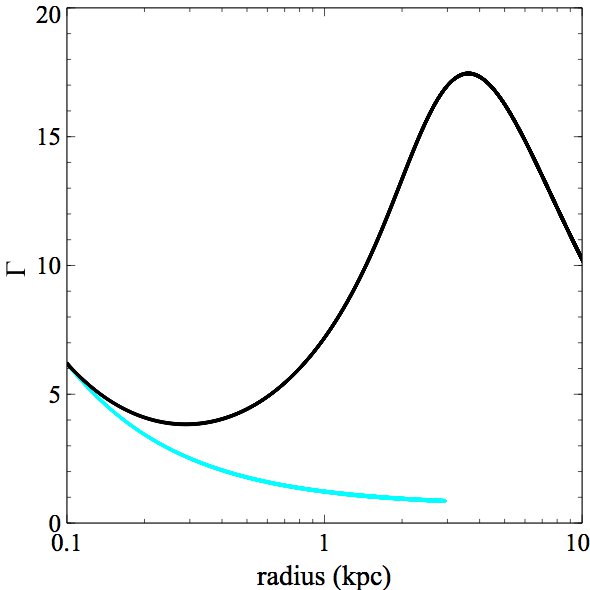} 
\caption{\small
Effective Eddington ratio as a function of radius: fixed-mass shell (black) and isothermal shell (cyan). 
The central luminosity is $L = 10^{47}$erg/s, the fixed-mass and isothermal shells are initially matched to $N_{\rm sh,0} = 10^{24} cm^{-2}$ (for $R_0 =$100 pc). We note that the IR transparency radius is located at a smaller distance for the fixed-mass shell ($R_{IR} \sim 0.3$ kpc) than for the isothermal shell ($R_{IR} \sim 0.8$ kpc). 
}
\label{Fig_ISO_Gamma_r_comp}
\end{center}
\end{figure}

In Figure \ref{Fig_ISO_Gamma_r_comp}, we compare graphically the radial dependence of the Eddington ratio for fixed-mass and isothermal shell models.
As fiducial parameters, we take a central luminosity of $L = 10^{47}$erg/s, characteristic of hyper-luminous dusty quasars, and assume an initial column density corresponding to the Compton-thick limit ($N_{sh,0} = 10^{24} cm^{-2}$).  
In the single scattering regime, we observe that the Eddington ratio increases with radius for the fixed-mass shell, while this is not the case for the isothermal shell. Although identical in the IR-optically thick (and UV-optically thin) regimes, there is a clear divergence in the single scattering regime, which is responsible for the different dynamics of the outflowing shell.

\begin{figure}
\centering
\textbf{\quad \quad Velocity vs. radius} 
\begin{center}
\includegraphics[angle=0,width=0.4\textwidth]{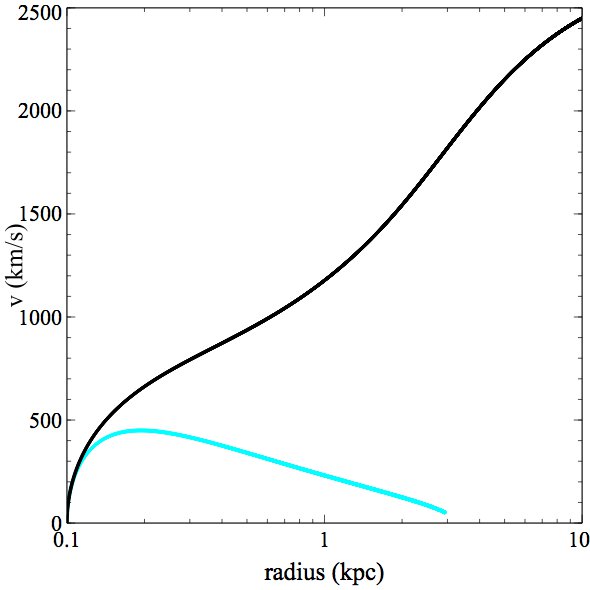} 
\caption{\small
Comparison of radial velocity profiles: fixed-mass shell (black) and isothermal shell (cyan). Same parameters as in Fig.\ref{Fig_ISO_Gamma_r_comp}.}
\label{Fig_v_r_comp}
\end{center}
\end{figure} 

This can be explicitly seen by considering the effective acceleration, which is a crucial parameter governing the shell dynamics. 
In the case of fixed-mass shells, the effective acceleration is given by
\begin{equation}
a = \frac{dv}{dt} = \frac{L}{c M_{\rm sh}} (1 + \tau_{\rm IR} - e^{-\tau_{\rm UV}} ) - \frac{GM(r)}{r^2}
\end{equation}
While $a_{IR}$ and $a_{UV}$ are the same for fixed-mass and isothermal shells, the net acceleration in the single scattering regime is constant for fixed-mass shells ($a_{SS,fix} = \frac{L}{c M_{\rm sh}}$), while it decreases with radius for isothermal shells ($a_{SS,iso} = \frac{L}{c 4 \pi m_p R_0^2 n_0 r}$). 
Moreover, in the case of isothermal shell models, an additional factor ($-\frac{v^2}{r}$) is present. 
These differences indicate that isothermal shells are less efficiently accelerated than fixed-mass shells, with major implications on the radial velocity profiles and obscuration properties (see below). This further underlines the importance of the single scattering regime. 

Figure \ref{Fig_v_r_comp} shows the comparison of the radial velocity profiles for fixed-mass and isothermal shell models. We clearly see a great difference in the shell dynamics, due to the above mentioned physical differences in the single scattering limit. The fixed-mass shell is efficiently accelerated to high velocities, exceeding $v \gtrsim 1000$ km/s on $\gtrsim$kpc scales; whereas the isothermal shell has much lower velocities and is decelerated due to the continuous sweeping-up of ambient material.


\section{Obscuration in dusty outflows }
\label{Sect_Obscuration}

We now consider the obscuration properties of the dusty shells driven by radiative feedback. 
In the following, we assume a Navarro-Frenk-White (NFW) potential given by
\begin{equation}
\rho(r) = \frac{\rho_\mathrm{c} \delta_\mathrm{c}}{(r/r_\mathrm{s})(1+r/r_\mathrm{s})^2} \, , 
\end{equation} 
where $\rho_\mathrm{c} = \frac{3 H^2}{8 \pi G}$ is the critical density, $\delta_\mathrm{c}$ is a characteristic density, and $r_\mathrm{s}$ is a scale radius.

The obscuration of the source is characterised by the shell column density, $N_{sh}(r)$. 
The evolution of the column density can also be expressed in terms of the extinction. One can assume a linear relation between optical extinction and hydrogen column density, of the form: $N_{\rm H} (\mathrm{cm^{-2}}) = (2.21 \pm 0.09) \times 10^{21} A_{\rm V} (\mathrm{mag})$ \citep{Guver_Ozel_2009}, or the refined empirical relation $N_{\rm H} = (2.87 \pm 0.12) \times 10^{21} A_{\rm V} \,\mathrm{cm^{-2}}$ \citep{Foight_et_2016}. 


In addition to the fixed-mass and isothermal shell models, we consider expanding shells with radial density profiles declining more steeply than isothermal at large radii:
\begin{equation}
n(r) \propto \frac{1}{r^2 (r+r_a)^{\gamma}}
\end{equation}
where $r_a$ is a scale radius and $\gamma > 0$.  
We analyse the cases of expanding shells with $r_a \sim 1$kpc and $\gamma = 1/4, 1/2$ (note that in this formalism the isothermal shell would correspond to the case $\gamma = 0$).

\begin{figure}
\centering
\textbf{\quad \quad Column density vs. velocity} 
\begin{center}
\includegraphics[angle=0,width=0.4\textwidth]{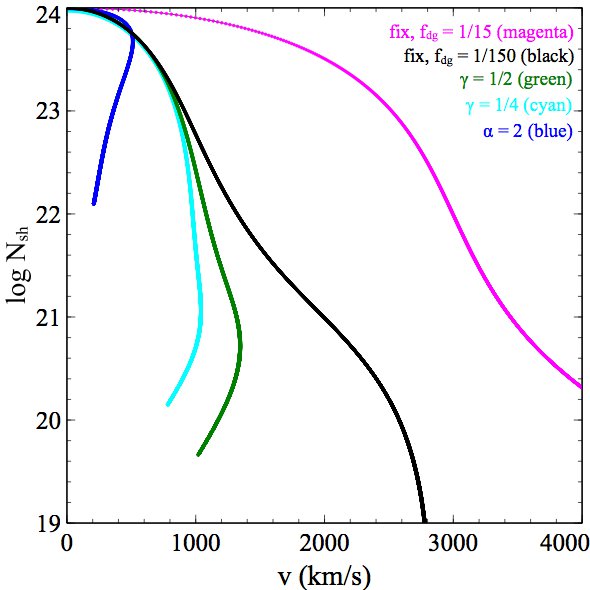} 
\caption{\small
Shell column density as a function of velocity for different shell models ($L = 10^{47}$erg/s, $N_{\rm sh,0} = 10^{24} cm^{-2}$): fixed-mass shell with $f_{\rm dg} =1/150$ (black), isothermal shell with $\alpha = 2$ (blue), expanding shell with $\gamma = 1/4$ (cyan), expanding shell with $\gamma = 1/2$ (green), fixed-mass shell with enhanced $f_{\rm dg} =1/15$ (magenta). 
}
\label{Fig_Nsh_v_5}
\end{center}
\end{figure}

In Fig. \ref{Fig_Nsh_v_5}, we plot the shell column density as a function of velocity for the different shell mass configurations. For given initial conditions, corresponding to the Compton-thick limit ($N_{\rm sh,0} \sim 10^{24} cm^{-2}$), the shell mass is $M_{\rm sh} = 9 \times 10^8 M_{\odot}$ in the fixed-mass shell model, comparable to the range of shell masses ($M_{\rm sh}  \sim 4 \times 10^8 M_{\odot} - 4 \times 10^9 M_{\odot}$) assumed in e.g. \citet{Thompson_et_2015}. For the expanding shell models with $\gamma = 1/2$ and $\gamma = 1/4$, the shell mass increases with radius as $\propto \sqrt{r+r_a}$ and $\propto (r+r_a)^{3/4}$ respectively, reaching typical values of $M_{\rm sh} \sim (1- 5) \times 10^9  M_{\odot}$ in the (1-10) kpc range. 
Starting from the same initial column density, we see that the subsequent shell evolutions are quite different: the fixed-mass shell (black curve) is accelerated to high velocities, and the associated column density rapidly falls off; whereas the isothermal shell (blue curve) has a much lower speed, with the corresponding column density decreasing much more slowly. The expanding shells with $\gamma = 1/4$ and $\gamma =1/2$ (cyan and green curves) present an intermediate behaviour between the two extremes. 
Overall, we see that the obscuration is rapidly dissipated in the case of fixed-mass shells, while it is more persistent for expanding shells sweeping up ambient material.

In the same Figure \ref{Fig_Nsh_v_5} we also show the effect of varying the dust-to-gas ratio. 
Recent ALMA observations of cold dust in dust-reddened quasars indicate a range of dust-to-gas ratios, from $f_{\rm dg} \sim 1/100$ to $f_{\rm dg} \sim 1/30$ \citep{Banerji_et_2017}. We observe that for an enhanced $f_{\rm dg} = 1/15$, the fixed-mass shell can easily attain velocities of several thousand km/s (magenta curve). In fact, an increase in the dust-to-gas fraction leads to higher shell velocities (as $\Gamma_{\rm IR} \propto \kappa_{\rm IR} \propto f_{\rm dg}$), and lower column densities at a given time. Thus variations in the dust-to-gas ratio can have a considerable impact on the shell dynamics.

\begin{figure}
\centering
\textbf{\quad \quad Shell extinction vs. time}
\begin{center}
\includegraphics[angle=0,width=0.4\textwidth]{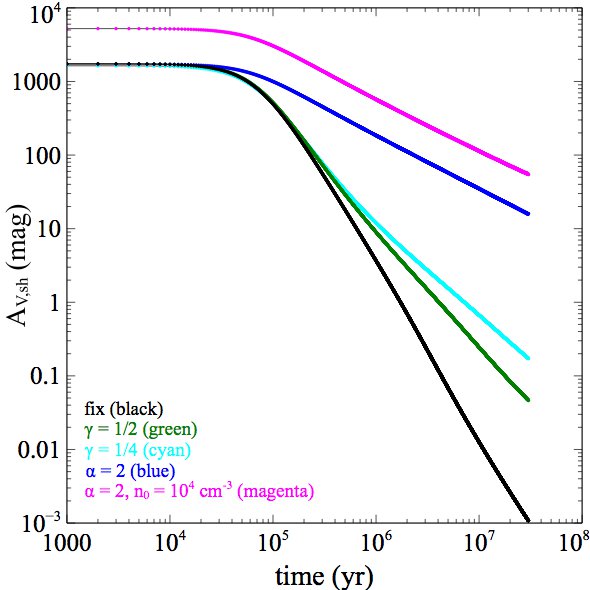} 
\caption{\small 
Shell extinction as a function of time for different shell models ($L = 10^{47}$erg/s, $N_{\rm sh,0} = 10^{24} cm^{-2}$, $f_{\rm dg} = 1/30$): fixed-mass shell (black), isothermal shell with $\alpha = 2$ (blue), expanding shell with $\gamma = 1/4$ (cyan), expanding shell with $\gamma = 1/2$ (green), isothermal shell with enhanced $n_0 = 10^4 cm^{-3}$ (magenta). 
}
\label{Fig_AV_t_4_bis}
\end{center}
\end{figure}

Figure \ref{Fig_AV_t_4_bis} shows the shell extinction as a function of time for the different shell mass configurations, with the dust-to-gas ratio fixed to an intermediate value ($f_{\rm dg} = 1/30$). For the expanding shell models, we consider density values in the range $n_0 \sim (10^2 - 10^4) \mathrm{cm^{-3}}$, which correspond to typical densities expected in the inner regions of dense starbursts \citep{Murray_et_2010}. 
For a given initial column density, we see that the extinction falls off more rapidly for fixed-mass shells compared to the expanding shells. In particular, the extinction can be orders of magnitude larger for isothermal shells compared to fixed-mass shells. 
At a given time, the extinction is larger for expanding shells with flatter density profiles (i.e. smaller $\gamma$), and we observe that this divergence increases with time. 

In Fig. \ref{Fig_AV_t_4_bis}, we also show that an enhancement in the external density naturally implies higher obscuration, for the case of the isothermal shell (magenta curve). In fact, as the external density increases, the amount of swept-up material increases, and since the effective acceleration decreases in the single scattering regime (as $a_{SS,iso} \propto \frac{1}{n_0}$), the outflowing shell is less easily accelerated, leaving a higher obscuring column. But we note that further increasing the external density (to values of $n_0 > 10^5 cm^{-3}$) does not change much the shell dynamics, since the effective acceleration in the IR-optically thick regime ($a_{IR} = \frac{\kappa_{\rm IR} L}{4 \pi c r^2}$) is independent of $n_0$.

From Figures \ref{Fig_Nsh_v_5} and \ref{Fig_AV_t_4_bis}, we observe that high-luminosity ($L \sim 10^{47}$erg/s) sources can have a considerable dust-obscured phase. 
As a general trend, the obscuration is more important and seems to last longer for expanding shells (sweeping up ambient matter) with flatter density distributions. We suggest that such dusty shell models may describe the populations of luminous, dust-obscured quasars. 
Interestingly, a number of dust-reddened quasars show direct signs of outflowing gas and some evidence for strong outflows with velocities of $v \sim 800-1000$ km/s \citep{Banerji_et_2012}. 
Assuming that the dusty quasars can be modelled by expanding shells with $\gamma = (1/4 -1/2)$, we see that dusty columns with log $N_{\rm sh} \sim 21-22$ should typically be outflowing at speeds of $v \sim 1000$ km/s, in agreement with observations (Fig. \ref{Fig_Nsh_v_5}).

On the other hand, extreme outflows with velocities of $v \gtrsim 3000$ km/s on galactic scales have been discovered in a subset of ERQs \citep{Zakamska_et_2016}. Such high-velocity outflows cannot be adequately described by any expanding shell model, but could be interpreted in terms of fixed-mass shells with substantial amounts of dust. 
In Figure \ref{Fig_Nsh_v_5}, we see that the fixed-mass shell with enhanced dust-to-gas ratio (magenta curve) can attain velocities of several thousand km/s on large scales, comparable to the extreme outflows detected in ERQs. 
Among the ERQ population, a particularly high fraction of red quasars is also found to present BAL-like features \citep{Hamann_et_2017}. 
In our picture, BAL-like outflows and other forms of extreme outflows could be accounted for by fixed-mass shell models coupled with a high dust content. In fact, the observed outflows are found to be dusty, with a clear effect of dust extinction seen in the shape of the emission line profiles \citep{Zakamska_et_2016}. 

\begin{figure}
\centering
\textbf{\quad \quad Column density vs. velocity}
\begin{center}
\includegraphics[angle=0,width=0.4\textwidth]{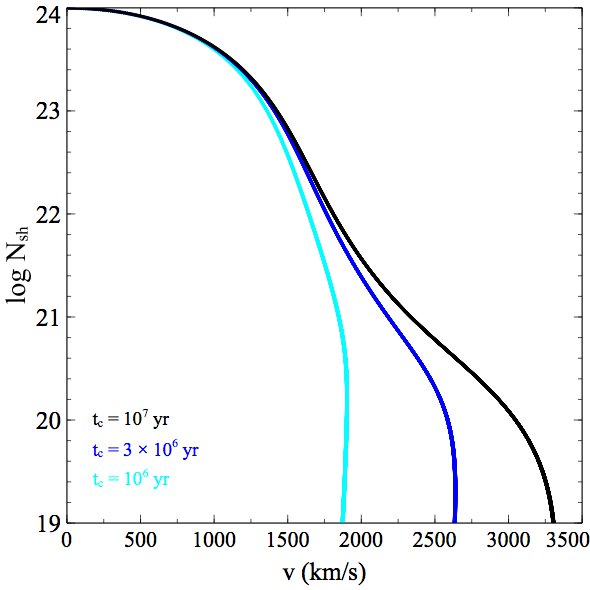} 
\caption{\small 
Shell column density as a function of velocity for fixed-mass shells with exponential luminosity decay: $L_0 = 10^{47}$erg/s, $t_c = 10^7$ yr (black), $t_c = 3 \times 10^6$ yr (blue), $t_c = 10^6$ yr (cyan). 
}
\label{Fig_Nsh_v_expLdec_3}
\end{center}
\end{figure}

\begin{figure}
\centering
\textbf{\quad \quad Shell extinction vs. time}
\begin{center}
\includegraphics[angle=0,width=0.4\textwidth]{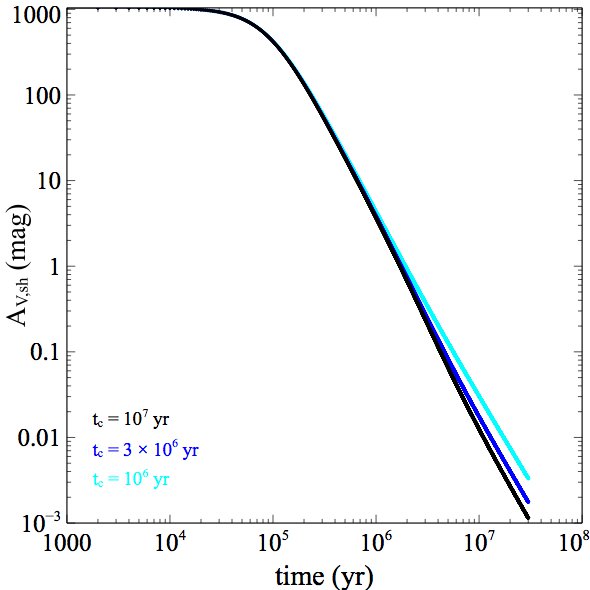} 
\caption{\small 
Shell extinction as a function of time for fixed-mass shells with exponential luminosity decay. Parameters as in Fig. \ref{Fig_Nsh_v_expLdec_3}. 
}
\label{Fig_AV_t_expLdec_3}
\end{center}
\end{figure}

A further aspect to consider is the temporal variability of the central luminosity output, which can affect the shell dynamics and obscuration properties. As an example, we examine the case of an exponential decay in luminosity: $L(t) = L_0 \exp^{-t/t_c}$, where $L_0$ is the initial luminosity and $t_c$ is a characteristic timescale. In Figure \ref{Fig_Nsh_v_expLdec_3} we show the effects of such luminosity variations in the case of fixed-mass shell models (with $L_0 = 10^{47}$erg/s and $f_{\rm dg} = 1/50$). 
We observe that the shell velocity is reduced, and the column density decreases more slowly, for exponential decays with shorter characteristic timescales. 
The corresponding shell extinction as a function of time is shown in Figure \ref{Fig_AV_t_expLdec_3}. Within the fixed-mass shell model, only a small difference is seen at later times. 

\begin{figure}
\centering
\textbf{\quad \quad Shell extinction vs. time}
\begin{center}
\includegraphics[angle=0,width=0.4\textwidth]{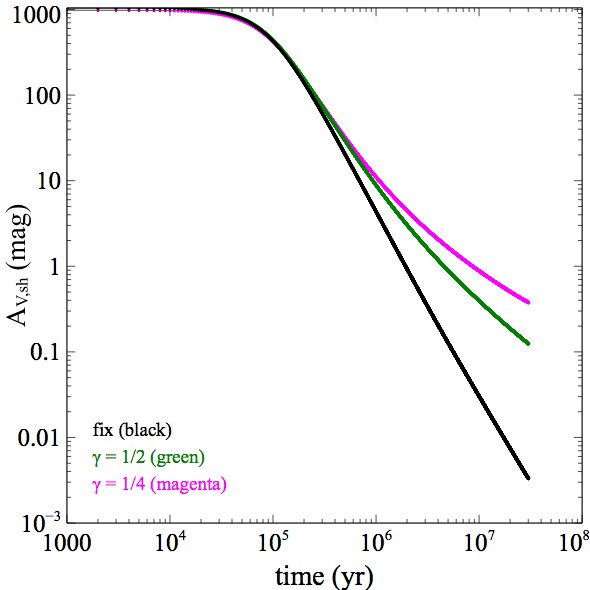} 
\caption{\small 
Shell extinction as a function of time for different shell models with exponential luminosity decay ($t_c = 10^6$yr): fixed-mass shell (black), expanding shell with $\gamma = 1/2$ (green), expanding shell with $\gamma = 1/4$ (magenta).
}
\label{Fig_AV_t_expLdec_comp}
\end{center}
\end{figure}

Figure \ref{Fig_AV_t_expLdec_comp} shows the temporal evolution of the shell extinction for different shell mass configurations. 
We observe that expanding shells have longer obscured phases, in particular for flatter ambient density distributions. For instance, the obscured phase with $A_{\rm V,sh}$ in the (1-10) mag range is rather short ($\sim 10^6$yr) for fixed-mass shells, while it can be longer ($\sim 5 \times 10^6$yr) for expanding shells. 
Observational studies of AGN variability indicate that significant luminosity variations can occur on a range of timescales \citep[][and references therein]{Hickox_et_2014}. 
Therefore, both the ambient matter distribution and the AGN activity timescale can affect the obscuration properties of the source. 
We have previously argued that luminosity variations, due to the variable nature of the central AGN, can be important in interpreting the observed outflow characteristics \citep{Ishibashi_Fabian_2015}.

\section{Discussion}

Dusty quasars are thought to represent an evolutionary phase in the lifetime of galaxies, transitioning from dust-enshrouded starbursts to unobscured luminous QSOs. 
In Section \ref{Sect_Obscuration}, we have seen that the obscuration is more pronounced, and the obscured phase lasts longer, in the case of expanding shells sweeping up ambient material. 
For instance, for expanding shell models with $\gamma = (1/4-1/2)$, a typical obscured phase with $A_{\rm V} \sim (1-10)$ mag corresponds to timescales of a few $\sim 10^6$ yr (Fig. \ref{Fig_AV_t_4_bis}). This may be consistent with the interpretation that the dusty quasars represent a short-lived blow-out phase in the evolution of galaxies \citep{Banerji_et_2015}.  
At the highest luminosities, observations suggest that the number density of dust-obscured AGNs is comparable to, or even exceed that of, unobscured quasars. 
In general, for a given shell mass configuration, an increase in the luminosity leads to a faster removal of the obscuring column and hence lower obscuration (since the effective Eddington ratio increases with $L$ in all three regimes, cf Eq. \ref{Gamma_IR_UV} - \ref{Gamma_SS_iso}).  
However, if we compare different shell mass configurations, we note that even assuming higher luminosities for expanding shells, the obscuration can still be larger than in the case of fixed-mass shells (for the observed range of $A_{\rm V} \sim$few mag).
This may account for the large number of dusty quasars observed at high luminosities \citep{Banerji_et_2015}. 

Within the evolutionary picture, the more heavily obscured sources (with $A_{\rm V}$$\sim$50 mag), such as Hot DOGs, could correspond to earlier temporal stages in the evolutionary sequence.  At early times, the central source is still enshrouded, and the dusty outflow is just starting to emerge from the dense cocoon. 
An observational example is given by the most luminous Hot DOG W2246Ð0526, which is close to the point of blowing out its dusty cocoon in a large-scale, isotropic outflow \citep{Diaz-Santos_et_2016}. 
Alternatively, the ambient density distribution may be different, and Hot DOGs may be better described by expanding shells with flatter slopes (e.g. isothermal shells or even $\alpha < 2$ shells). Indeed, a flatter slope in the ambient density distribution implies a steeper radial dependence for the swept-up material, leading to strong shell deceleration and prolonged obscuration. 

Therefore, in addition to the temporal sequence, differences in the surrounding density distribution may also contribute to the diverse observational appearance. The different shell mass configurations are essentially determined by the different physical conditions of the ambient medium. If large amounts of gas are present in the inner regions, the outflowing shell continues to sweep up mass as it expands; while the fixed-mass shell approximation may be more appropriate in lower density environments.  
In the latter case, the partial depletion of gas could be in part attributed to previous feedback-driven outflow events, which in turn may facilitate the subsequent outflow episodes. 

A complex interplay between luminosity and obscuration may arise at different stages in the evolutionary sequence. 
At early times, e.g. following a merger event, large quantities of gas are funnelled towards the centre, leading to both high luminosity and high extinction. This suggests a positive coupling, whereby the accreting material is also responsible for significant obscuration.
Once the IR optical depth becomes large enough, the system enters in the IR-optically thick regime, and the effective Eddington ratio $\Gamma_{\rm IR}$ becomes independent of the column density. This implies that even dense material can be disrupted, allowing the potential development of a radiation pressure-driven outflow. Since radiative feedback directly acts on the same dusty gas, which also provides obscuration, a causal link between dust obscuration and blow-out may be expected. 
In this picture, large amounts of dust imply heavy obscuration, but also powerful feedback. 
In subsequent stages, an increase in the central luminosity leads to higher Eddington ratios, and hence powerful outflows, which rapidly clear the obscuring dusty gas. In these phases, one may expect an anti-correlation between luminosity and obscuration due to radiative feedback. We have previously discussed how such AGN radiative feedback may provide a natural physical interpretation of the observed evolutionary sequence \citep{Ishibashi_Fabian_2016b}.  

The coupling between radiation field and dusty gas in extreme environments (e.g. ULIRG) has been investigated via numerical simulations. Early results, based on the flux-limited diffusion (FLD) approximation, suggested that the rate of momentum transfer is considerably reduced due to the development of radiative Rayleigh-Taylor instabilities \citep{Krumholz_Thompson_2013}. But subsequent simulations, using the more accurate variable Eddington tensor (VET) approximation, indicate that dusty gas can be continuously accelerated, despite the development of instabilities in the flow \citep{Davis_et_2014}.
The most recent study re-examining the issue (with radiation hydrodynamic simulations based on the VET method) indicates that the momentum transfer can be significantly amplified with respect to the single scattering limit, confirming that radiation pressure on dust can be an important mechanism for driving large-scale outflows \citep{Zhang_Davis_2016}.


\section*{Acknowledgements }

WI acknowledges the Physik-Institut for hospitality. 
MB acknowledges funding from the UK Science and Technology Facilities Council (STFC) via an Ernest Rutherford Fellowship. 
ACF acknowledges ERC Advanced Grant 340442.

  
\bibliographystyle{mn2e}
\bibliography{biblio.bib}


\label{lastpage}

\end{document}